\documentclass[%
 reprint,
 amsmath,amssymb,
 aps,
prb,
showkeys,
]{revtex4-2}

\usepackage{graphicx}
\usepackage[export]{adjustbox}
\usepackage{dcolumn}
\usepackage[ruled]{algorithm2e}
\usepackage{bm}
\usepackage{enumitem}
\usepackage{xcolor}
\begin{document}

\preprint{APS/123-QED}

\title{The Levy-Lieb embedding of density functional theory and its Quantum Kernel: Illustration for the Hubbard Dimer using near-term quantum algorithms}

\author{C. D. Pemmaraju}
\email{dasc@ibm.com}
\affiliation{IBM, San Jose, Califonia, USA}
%


\author{Amol Deshmukh}
\affiliation{
IBM Quantum, IBM T.J. Watson Research Center, Yorktown Heights, NY, USA
}%


\date{\today}

\begin{abstract}
The constrained-search formulation of Levy and Lieb provides a concrete mapping from $N$-representable densities to the space of $N$-particle wavefunctions and explicitly defines the universal functional of density functional theory. We numerically implement the Levy-Lieb procedure for a paradigmatic lattice system, the Hubbard dimer, using a modified variational quantum eigensolver (VQE) approach. We demonstrate density variational minimization using the resulting hybrid quantum-classical scheme featuring real-time computation of the Levy-Lieb functional along the search trajectory. We further illustrate a fidelity based quantum kernel associated with the density to pure-state embedding implied by the Levy-Lieb procedure and employ the kernel for learning observable functionals of the density. We study the kernel's ability to generalize with high accuracy through numerical experiments on the Hubbard dimer. 

\end{abstract}

\keywords{Quantum Computing, Quantum Kernel, Machine Learning, Variational Quantum Eigensolver (VQE), Density Functional Theory (DFT)}
\maketitle


\section{\label{sec:intro}Introduction:\protect}

Density Functional Theory (DFT) ~\cite{Hohenberg1964, Kohn1965} is presently the dominant paradigm for material-specific electronic structure simulations in computational materials science~\cite{VanNoorden2014}. Originally proposed by Hohenberg and Kohn (HK)~\cite{Hohenberg1964}, DFT establishes the one-body density as the basic variable in interacting many-body problems with fixed interaction strength subject to time-independent spatially local external potentials~\cite{Parr1989}. The electronic Hamiltonian in the Born-Oppenheimer approximation is the most prominent example of such a setup. Given a problem in this class, all ground state observables can in principle be represented as explicit functionals of the ground state one-body density. However, while the density is informationally adequate, deriving explicit density functionals that are universally accurate is a hard problem in the same complexity class as computing generic ground states~\cite{Schuch2009}. The practical success of DFT instead stems from the development of extremely low cost yet usefully accurate~\cite{Zhang2018} \textit{approximate models} for the quantum mechanical correlation energy in the form of highly compact density functionals~\cite{Mardirossian2017} circumventing the need to represent many-electron wavefunctions explicitly on classical computers at simulation time. Benefits of the compactness of a density functional representation can also extend to quantum systems where efficient classical heuristics for the wavefunction exist~\cite{Li2016, Capelle2013}. In the modern context the encodings implied by DFT may be viewed as simply the most condensed description within a hierarchy of functional theories based on reduced density matrices~\cite{Schuch2009, Ludena2013}.

With the advent of quantum computers the prospect of representing many-electron states on quantum hardware has important implications for DFT and other functional theories~\cite{Schuch2009, Gaitan2009, Baker2020, Hatcher2019, Seanjean2022}. The classical intractability of interacting many-electron wavefunctions partly provided the motivation for functional theories~\cite{Kohn1999} but at the same time also constrained the development of approximate functionals~\cite{Capelle2013}. For instance the study of exact forms of DFT has to date been largely restricted to small model systems where systematic comparisons with wavefunction calculations are possible~\cite{Capelle2013} and pure density functionals for a wider variety of observables are known in model systems than are available for deployment in realistic simulations of matter~\cite{Capelle2013, Li2016, Carrascal2015, Carrascal2018}. Furthermore, the relationship between approximate functionals and underlying many-electron states is often obscured in practice even though the mapping can be recovered \textit{ex post facto} at significant computational cost~\cite{Coe2010}. Finally and most relevant to this work is the fact that outside of a few pioneering efforts~\cite{Cioslowski1988, Cioslowski1989, Cioslowski1991, Capelle2003, DAmico2011, Mori-Sanchez2018} the explicit density to wavefunction mappings implied by DFT~\cite{Levy1979, Levy1982, Lieb1983, Levy1985, Zhao1993} are not widely pursued in numerical studies on classical computers and one typically encounters density functionals only after they are approximated either using formal, empirical or machine learning methods~\cite{Mardirossian2017, vonLilienfeld2020}.  The availability of quantum devices may alleviate this constraint in future. Therefore in this work we calculate the density to wavefunction mapping~\cite{Levy1985, Zhao1993, Eschrig2003, Cioslowski1988, Cioslowski1989, Cioslowski1991, Mori-Sanchez2018},  as formulated by Levy~\cite{Levy1979, Levy1982} and Lieb~\cite{Lieb1983} using a density-constrained variational quantum eigensolver (VQE)~\cite{Peruzzo2014, Tilly2021} scheme and demonstrate density variational minimization to identify the ground state of a paradigmatic fermion lattice problem, the Hubbard dimer~\cite{Carrascal2015, Carrascal2018}. Furthermore, we make use of the density-wavefunction map to define a quantum kernel~\cite{Havlicek2019, Schuld2019, Schuld2021, Kubler2021} which we call the Levy-Lieb quantum kernel and use it to learn observable functionals in such a way that the relationship to the underlying state space is readily apparent. Our work contrasts with previous research efforts which explored proposals for the popular Kohn-Sham DFT~\cite{Kohn1965} in both near-term and fault-tolerant settings~\cite{Gaitan2009, Baker2020, Hatcher2019, Seanjean2022}. The rest of the article is organized as follows. In section~\ref{sec:levylieb}, we introduce the Levy-Lieb procedure which defines the density to wavefunction map and in section~\ref{sec: dcvqe} we outline a concrete VQE based implementation of the same for lattice systems. Section ~\ref{sec: hubdim} discusses results of an explicit density variational search for the ground state of the Hubbard dimer. In section~\ref{sec:llqk} we introduce a fidelity based Levy-Lieb quantum kernel (LLQK) and illustrate its use in machine learning density functionals.  We summarize our conclusions in section~\ref{sec: summary}.
\begin{figure*}[htbp]
\includegraphics[angle=0, width=\textwidth]{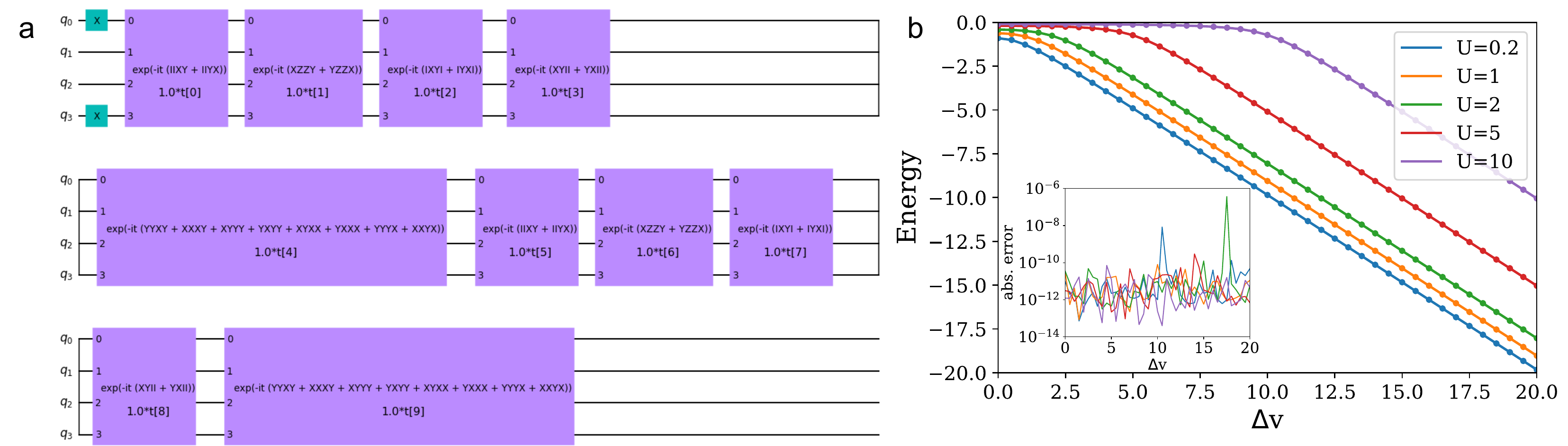}
\caption{\label{fig:circ} (a) Qiskit~\cite{Qiskit} circuit diagram for the generalized unitary coupled cluster ansatz used to prepare parameterized quantum states for the asymmetric Hubbard dimer studied in this work. (b) For different values of the $U$ parameter, a comparison of VQE (filled circles) and exact diagonalization (solid line) results for the ground state energy of the Hubbard dimer as a function of the potential asymmetry $\Delta v$. The inset shows the absolute error in the VQE energy relative to the exact diagonalization reference on a logarithmic scale.}
\end{figure*}
\section{\label{sec:levylieb} The Levy-Lieb mapping}
Consider an interacting $N$-particle system described by a Hamiltonian operator of the form
\begin{equation}\label{eq:Ham}
\hat{H} = \hat{T} + \hat{W} + \hat{v}    
\end{equation}
where $\hat{T},\hat{W}$ are the kinetic and inter-particle interaction operators and the external potential $\hat{v}$ is a local one-body operator. 
Following Eschrig~\cite{Eschrig2003}, we define the set of $N$-representable densities as
\begin{equation}
\mathcal{J}_N \equiv \{n | n \ge 0, \nabla n^{1/2} \in \boldsymbol{L}^2, \int dx~n=N \}
\end{equation}
with $\boldsymbol{L}^2$ denoting the space of square-integrable functions, and the set of $N$-particle wavefunctions as
\begin{eqnarray}
\mathcal{W}_N \equiv \{ \Psi |&& \Psi(x_1,..,x_N) \mathrm{(anti)symmetric}, \langle\Psi|\Psi\rangle = 1, \nonumber\\
&& \langle\nabla_i\Psi|\nabla_i\Psi\rangle < \infty~\mathrm{for}~i=1..N  \}
\end{eqnarray}
The Levy-Lieb density functional $F_\mathrm{LL}[n]$ is then defined by 
\begin{equation}\label{eq:llf}
F_\mathrm{LL}[n]\equiv\mathrm{inf}\{\langle\Psi|\hat{T}+\hat{W}|\Psi\rangle | \Psi\rightsquigarrow~n, \Psi \in \mathcal{W}_N\}, n \in \mathcal{J}_N 
\end{equation}
i.e., given a density $n \in \mathcal{J}_N$, the Levy-Lieb procedure involves searching for the infimum of $\langle\Psi|\hat{T}+\hat{W}|\Psi\rangle$ over the restricted subset of wavefunctions $\Psi \in \mathcal{W}_N$ constrained to yield the chosen density $n$. We use the notation $\Psi \rightsquigarrow n$ to indicate $\Psi$ yields $n$ and define $\mathcal{W}^n_N \equiv \{ \Psi \in \mathcal{W}_N | \Psi \rightsquigarrow n \}$. Note that the above procedure does not involve the external potential $\hat{v}$ and at the end of the search process we have access to both the value of $F_\mathrm{LL}[n]$ and a $\Psi \rightsquigarrow n$. For general $n \in \mathcal{J}_N$ the obtained $\Psi$ belongs to a sub-manifold of states on $\mathcal{W}^n_N$ with the same expectation $\langle\Psi|\hat{T}+\hat{W}|\Psi\rangle$~\cite{Eschrig2003} but in the absence of such degeneracies on $\mathcal{W}^n_N$, $n$ uniquely determines $\Psi$~\cite{Capelle2007}. In particular, on the subset of densities that correspond to non-degenerate ground states, the mapping is bijective due to the Hohenberg-Kohn theorem~\cite{Hohenberg1964, Eschrig2003, Capelle2007}. In all cases, $F_\mathrm{LL}[n]$ is well-defined. The density variational principle follows directly from the Levy-Lieb procedure. Consider the set $\mathcal{W}_N$ and define the equivalence relation
\begin{equation}
\Psi_i \sim \Psi_j \iff \Psi_i \rightsquigarrow n~\mathrm{and}~ \Psi_j \rightsquigarrow n
\end{equation}
This leads to a partitioning of $\mathcal{W}_N$ into disjoint subsets $\mathcal{W}^n_N$ labelled by the density $n$ as $\Psi_i \in \mathcal{W}^n_N~\mathrm{iff}~\Psi_i \rightsquigarrow n $. On a particular $\mathcal{W}^n_N$, consider minimizing the expectation of the Hamiltonian from equation~\ref{eq:Ham} with external potential $\hat{v}$
\begin{eqnarray}
E[\hat{v},n] &&=\mathrm{inf}\{ \langle\Psi|\hat{T}+\hat{W}+\hat{v}|\Psi\rangle, \Psi \in \mathcal{W}^n_N\} \\ \nonumber
&&=\mathrm{inf}\{ \langle\Psi|\hat{T}+\hat{W}|\Psi\rangle+ \int dx~v(x)n(x), \Psi \in \mathcal{W}^n_N\} \\ \nonumber
&&=\mathrm{inf}\{ \langle\Psi|\hat{T}+\hat{W}|\Psi\rangle, \Psi \in \mathcal{W}^n_N\}+ \int dx~v(x)n(x)\\ \nonumber
&&=F_\mathrm{LL}[n]+ \int dx~v(x)n(x)
\end{eqnarray}
Since $\mathcal{W}_N = \bigcup\limits_{n\in \mathcal{J}_N} \mathcal{W}^n_N $, for the ground state energy we have
\begin{eqnarray}\label{eq:dvp}
E[\hat{v}] && \equiv \mathrm{min}\{\langle\Psi|\hat{T}+\hat{W}+\hat{v}|\Psi\rangle, \Psi \in \mathcal{W}_N \} \nonumber \\
    && =\mathrm{min}\left\{ \mathrm{inf}\{\langle\Psi|\hat{T}+\hat{W}+\hat{v}|\Psi\rangle, \Psi \in \mathcal{W}^n_N \}, n \in \mathcal{J}_N \right\} \nonumber \\ 
    && = \mathrm{min}\{ F_\mathrm{LL}[n]+ \int dx~v(x)n(x) , n \in \mathcal{J}_N \} 
\end{eqnarray}
Thus the ground state energy $E[\hat{v}]$ can be obtained by minimizing $E[\hat{v},n] = F_\mathrm{LL}[n]+ \int dx~v(x)n(x)$ over $n \in \mathcal{J}_N$. It follows further that if $n_\mathrm{GS}$ is the ground-state density then through equation~\ref{eq:llf} on $\mathcal{W}^{n_\mathrm{GS}}_N$, it determines the ground-state manifold $\Psi_\mathrm{GS}$.
We note that the Hohenberg-Kohn (HK) functional is a restriction of $F_\mathrm{LL}[n]$ to the space of ground state densities~\cite{Eschrig2003}. 

As formulated by Cioslowski~\cite{Cioslowski1988, Cioslowski1989, Cioslowski1991} and previously illustrated by Mori-Sanchez~\textit{et al}~\cite{Mori-Sanchez2018} on classical computers, the minimization procedure of equation~\ref{eq:dvp} defines an \textit{exact} numerical density functional calculation for the ground state provided $F_\mathrm{LL}[n]$ and it's  derivative ${\delta}F_\mathrm{LL}[n]/{\delta}n$ are calculated explicitly through constrained search. In this work we calculate these quantities for discrete lattices by employing parameterized quantum circuits executed on a statevector simulator. In the following sections when discussing formal statements unrelated to specific implementation details we will indicate functionals as $A[n]$ using square brackets and italicized $n$ as above but when referring to discretized implementations where parametric differentiation is substituted for formal functional differentiation~\cite{Cioslowski1988, Cioslowski1989, Cioslowski1991, Gonis2016, Mori-Sanchez2018}, we will indicate functional dependence via $A(\mathbf{n})$ using parenthesis and bold $\mathbf{n}$ notation for the density.

\section{\label{sec: dcvqe} Density-constrained VQE}
Since first being proposed in 2014, the variational quantum eigensolver (VQE)~\cite{Peruzzo2014} has been widely adopted on near-term quantum computers as a versatile hybrid classical-quantum algorithm for optimization tasks employing parameterized ansatze. The incorporation of constraints into VQE variational optimization has also been discussed previously~\cite{Ryabinkin2019, Kuroiwa2021}. For a recent comprehensive review of VQE we refer readers to reference~\cite{Tilly2021}. For our purpose, since equation~\ref{eq:llf} implies a search over wavefunction space restricted to a specified density sector, we augment the usual VQE with a density-constraint. If the Hamiltonian takes a form where the potential term appears as a local one body operator $\hat{v} = \sum_{i} v_i \sum_{\sigma} \hat{c}^{\dagger}_{i\sigma}\hat{c}_{i\sigma}$ in second-quantization, then the one-body density operator of interest needed to specify the constraint is $\hat{n}_{i\sigma} =  \hat{c}^{\dagger}_{i\sigma}\hat{c}_{i\sigma}$ which is easily computed on any discrete index set $\{i\sigma\}$ enumerating basis modes $i$ with spin index $\sigma$. In the literature, this choice of relevant potential and density on a lattice goes by the name site-occupation functional theory~\cite{Capelle2013} but the same form also appears in connection with discretized real space grids~\cite{Mori-Sanchez2018}. Then for $N$ electrons on $M$ sites and parameterized ansatz states $|\psi(\theta)\rangle$ with real parameters $\theta$, the Levy-Lieb procedure consists of minimizing the expectation
\begin{equation}
F_\mathrm{LL}(\theta) = \langle \psi(\theta) | \hat{T} + \hat{W} | \psi(\theta) \rangle
\end{equation}
subject to the  constraints
\begin{equation}\label{eq:dcons}
\langle \psi(\theta)| \sum_{\sigma}\hat{c}^{\dagger}_{i\sigma}\hat{c}_{i\sigma} | \psi(\theta) \rangle =  n_i, i = 1..M 
\end{equation}
where $\{n_i, i=1..M\}$ are prescribed site occupations and additionally we have 
\begin{equation}
\begin{array}{l}
0 < n_i \le 2, \\
\sum\limits_{i=1}^{M} n_i =  N 
\end{array}
\end{equation}
The steps involved in density-constrained VQE (DC-VQE) are outlined in table~\ref{tab:dcvqe}. We expect based on the work of D'Amico et al~\cite{DAmico2011} that for $v$-representable densites, i.e., densities that correspond to $N$-particle ground states in some potential $v$, small errors in computing densities do not lead to large errors in the wavefunction as nearby densities get mapped to nearby wavefunctions in metric space and densities on lattices are $v$-representable under very reasonable assumptions~\cite{Kohn1983, Chayes1985}. Thus the constrained minimization process should be robust to small numerical errors in computing densities. 
\begin{table}[h]
\caption{\label{tab:dcvqe}%
Density-constrained VQE (DC-VQE)
}
\begin{ruledtabular}
\begin{tabular}{l}
1. Accept a vector of site occupations \\ 
$~\mathbf{n} \equiv  [n_i, i=1..M]$ such that $\left\{
\begin{array}{c}
0 \le n_i \le 2\\
\sum_i n_i = N
\end{array}\right\}$ \\
2. Construct the vector of site density difference operators \\
$\mathbf{\hat{D}} \equiv [ \sum_{\sigma} \hat{c}^{\dagger}_{i\sigma}\hat{c}_{i\sigma} - n_i, i=1..M]$ \\
3. Set circuit parameters $\theta \leftarrow \theta_{guess}$ to prepare \\ 
initial state $\Psi(\theta) = R(\theta)|0\rangle$ \\
4. For $\mathbf{d} \equiv \langle \Psi(\theta)|\mathbf{\hat{D}}|\Psi(\theta)\rangle$ measured on a quantum device,\\
update $\theta$ to minimize $||\mathbf{d}||$ using a classical optimizer. \\
For $\mathbf{n} \in \mathcal{J}_N$, $||\mathbf{d}|| < \epsilon$ at the minimum and yields a state \\ 
$\Psi(\theta) \in \mathcal{W}^\mathbf{n^{\prime}}_N, \mathbf{n^{\prime}} \in B_{\epsilon}(\mathbf{n})$ \\
5. With $F_\mathrm{LL}(\theta) \equiv \langle\Psi(\theta)|\hat{T}+\hat{W}|\Psi(\theta)\rangle$ and $\mathbf{d} = \langle \Psi(\theta)|\mathbf{\hat{D}}|\Psi(\theta)\rangle$ \\ 
both measured on a quantum device, update $\theta$ to minimize \\ 
$F_\mathrm{LL}(\theta)$ under the constraint  $||\mathbf{d}|| = 0$ using a constrained \\
classical optimizer.\\
6. At the optimum $\theta^*$ set $F_\mathrm{LL}(\mathbf{n}) = F_\mathrm{LL}(\theta^*)$ and optionally \\ save $\theta^*$. Since $||\mathbf{d}|| < \epsilon$ we have $\Psi(\theta^*)  \rightsquigarrow \mathbf{n^{\prime}}, \mathbf{n^{\prime}} \in B_{\epsilon}(\mathbf{n})$. 
\end{tabular}
\end{ruledtabular}
\end{table}

Once a subroutine for calculating $F_\mathrm{LL}(\mathbf{n})$ is available, density variational minimization for a given external potential can be implemented as outlined in table~\ref{tab:dvs} wherein, following earlier implementations for classical computers~\cite{Cioslowski1988, Cioslowski1989, Cioslowski1991, Mori-Sanchez2018}, we compute the derivative ${\delta}F_\mathrm{LL}(\mathbf{n})/{\delta}\mathbf{n}$ using finite differences while maintaining normalization of the density. The next section outlines a specific implementation of DC-VQE and density variational search for the case of the asymmetric Hubbard dimer. 

\begin{table}[h]
\caption{\label{tab:dvs}%
Density variational minimization
}
\begin{ruledtabular}
\begin{tabular}{l}
1. Accept a vector of on-site potentials $~\mathbf{v} \equiv  [v_i, i=1..M]$\\ 
to define the full Hamiltonian \\
2. Use either a non-interacting or mean-field solution to setup\\
guess site occupations $\mathbf{n}$ and ansatz parameters $\theta_{guess}$ \\
3. Update $\mathbf{n}$ to minimize $F_\mathrm{LL}(\mathbf{n}) + \mathbf{v.n}$  subject to the \\
constraints $\left\{
\begin{array}{c}
0 < n_i \le 2\\
\sum_i n_i = N
\end{array}\right\}$ using a classical constrained \\ 
optimizer that calls DC-VQE to evaluate $F_\mathrm{LL}(\mathbf{n})$\\
4. At the optimum $\mathbf{n^*}$ return the ground state energy \\ 
$E(\mathbf{v}) = F_\mathrm{LL}(\mathbf{n^*}) + \mathbf{v.{n^*}}$  
\end{tabular}
\end{ruledtabular}
\end{table}

\begin{figure}[htbp]
\includegraphics[width=0.5\textwidth]{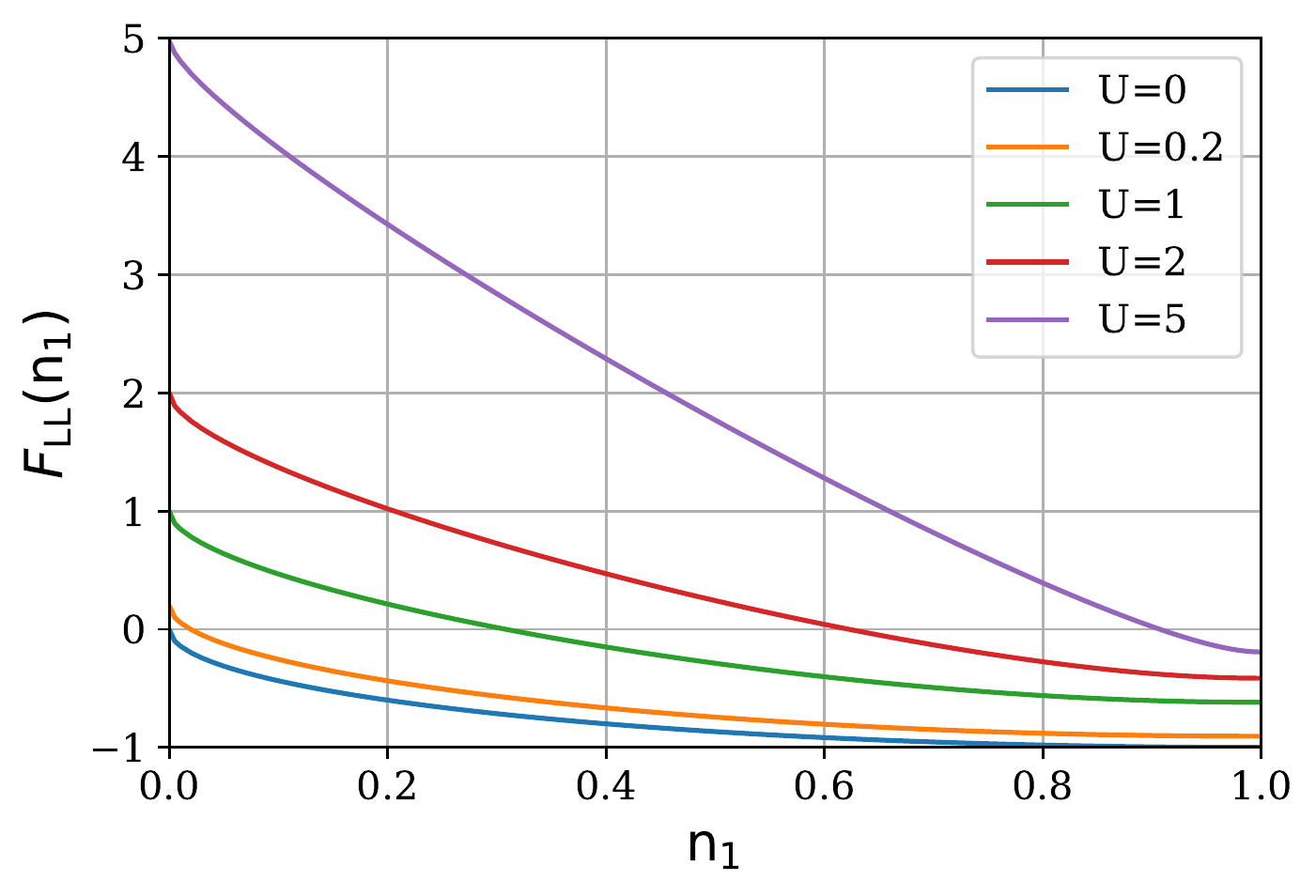}
\caption{\label{fig:lln} The Levy-Lieb functional for the asymmetric Hubbard dimer is shown for different values of the $U$ parameter.}
\end{figure}

\section{\label{sec: hubdim} Asymmetric Hubbard dimer}
The asymmetric Hubbard dimer has been explored extensively as a paradigmatic model system within lattice density functional theory~\cite{Carrascal2015} and time-dependent density functional theory~\cite{Carrascal2018} and serves as an ideal test case to explore near-term variational quantum algorithms and quantum machine learning in the context of DFT. The relevant fermion Hamiltonian 
\begin{equation}\label{eq:hubham}
\begin{aligned}
\hat{H} = & -t \sum_{\sigma=\uparrow,\downarrow} (c^{\dagger}_{1\sigma}c_{2\sigma} + h.c.)\\
& + U \sum_{i}\hat{n}_{i\uparrow}\hat{n}_{i\downarrow} + \sum_i v_i (\hat{n}_{i\uparrow}+\hat{n}_{i\downarrow})
\end{aligned}
\end{equation}
features a local multiplicative external potential term as the source of inhomogeneity in the model. It is further characterized by two dimensionless parameters $U/t$ and $\Delta v/t$ where $\Delta v = v_1-v_2$ is the potential asymmetry between the two dimer sites. To enable direct comparison with previous benchmark results~\cite{Carrascal2015, Carrascal2018} we fix $2t = 1$ and vary $U$, $\Delta v$ across different calculations. Additionally, since for the on-site occupations $\mathbf{n} = [n_1, n_2]$ we have $n_1 + n_2 = N$, a single parameter $n_1$ is chosen to specify the density distribution. We consider the case of $N=2$ particles in four site-localized spin orbitals and restrict our attention to the lowest energy spin-singlet ($S^2=0, S_z =0$) which is sufficient for analyzing ground state DFT of the dimer~\cite{Carrascal2015}. The above setup leads via the Jordan-Wigner mapping~\cite{Jordan1928} to a 4-qubit problem on a quantum computer. For this simple model we find that a generalized unitary coupled cluster~\cite{Lee2019} (UCC) wavefunction ansatz featuring single and double excitations along with two Trotter steps is sufficient to reproduce the exact-diagonalization result for the ground state energies over the parameter range of interest in conjunction with VQE~\cite{Peruzzo2014, Tilly2021} (See fig~\ref{fig:circ}). For a general discussion of specialized ansatze for lattice models in different dimensions we refer the readers to recent literature~\cite{Tilly2021}. In our study the UCC circuit ansatz which features ten parameters as shown in figure~\ref{fig:circ}(a) is set up using the Qiskit~\cite{Qiskit} framework and executed on a statevector simulator backend. VQE classical parameter optimization is performed using the L-BFGS-B~\cite{Zhu1997} optimizer as implemented in the SciPy~\cite{Virtanen2020} package. As shown in fig.~\ref{fig:circ}(b) for a range of $\left(U, \Delta v\right)$ parameters, energies of the dimer spin-singlet ground-state as obtained from VQE are within $10^{-6}$ of corresponding exact-diagonalization values and in good correspondence with figure $3$ of reference~\cite{Carrascal2015}. We then proceed to employ the same circuit ansatz to prepare states for evaluation within density-constrained VQE simulations of the dimer.

\begin{figure*}[htbp]
\includegraphics[angle=0, width=\textwidth]{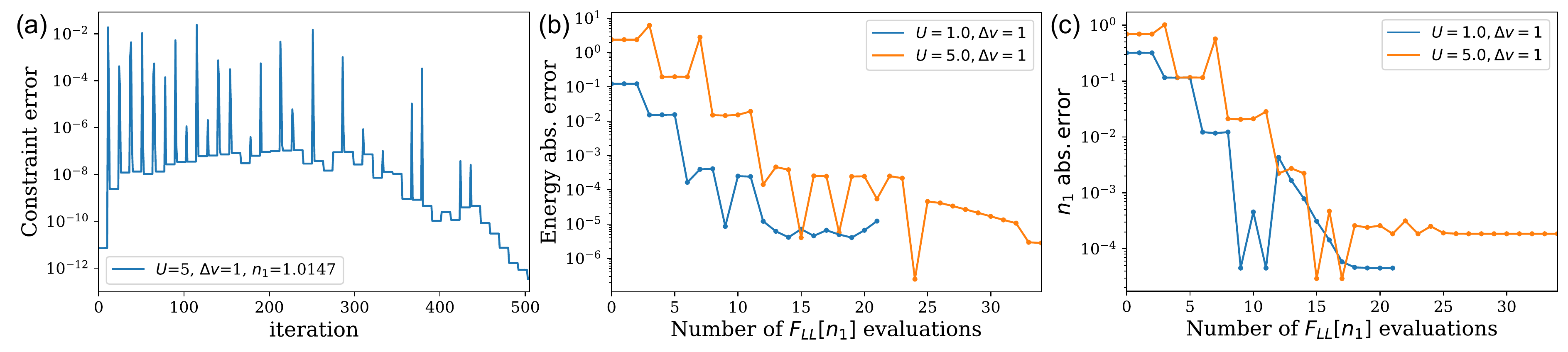}
\caption{\label{fig:dvs} (a) Evolution of the error in satisfying the density constraint of equation~\ref{eq:dcons} during the course of computing the Levy-Lieb functional via constrained optimization. (b) Absolute error relative to a VQE reference in the total energy during a density variational minimization for the ground state. (c) Absolute error relative to a VQE reference in the site occupation $n_1$ during a density variational minimization for the ground state.}
\end{figure*}

As outlined in section~\ref{sec: dcvqe} and table~\ref{tab:dcvqe}, density-constrained VQE (DC-VQE) is intended as a subroutine to calculate the Levy-Lieb functional $F_\mathrm{LL}(\mathbf{n})$ for a specified density $\mathbf{n}$ and makes no reference to the external potentials $\mathbf{v}$. Thus with the operator sum $\hat{T}+\hat{W}$ and a set of site-occupation difference operators $\mathbf{D}$ as defined in table~\ref{tab:dcvqe} specified, the UCC ansatz parameters are optimized with VQE under an occupation-preserving constraint (equation~\ref{eq:dcons}) to yield $F_\mathrm{LL}(\mathbf{n})$. We implement the constraint in practice by setting up the SLSQP~\cite{Kraft1988} constrained optimizer from SciPy~\cite{Virtanen2020} to request site-occupation measurements from the quantum simulator. For the dimer case, the DC-VQE simulated result for $F_\mathrm{LL}$ which is just a function of $n_1$ is plotted in figure~\ref{fig:lln} for different values of $U$. As noted previously by Carrascal et al~\cite{Carrascal2015}, there is no known analytical expression for $F_\mathrm{LL}(n_1)$ but it is a monotonic function on the interval $n_1 \in (0,1]$ and approaches a bound $U$ as $n_1 \rightarrow 0$. Furthermore since $\frac{dF_\mathrm{LL}(n_1)}{dn_1} = \frac{\Delta v}{2}$~\cite{Carrascal2015} its slope increases sharply as $n_1 \rightarrow 0$ but remains finite for physical potentials. It is apparent that the DC-VQE result for $F_\mathrm{LL}(n_1)$ exhibits the expected features.  

In connection with DC-VQE it is instructive to analyze the degree to which the occupation-preserving constraint of equation~\ref{eq:dcons} is respected during the optimization process for $F_\mathrm{LL}(n_1)$. This is shown in figure~\ref{fig:dvs}(a) where we plot the constraint vector magnitude $||\mathbf{d}||$ as defined in table~\ref{tab:dcvqe} along a typical optimization trajectory. We see that starting from a small value of $||\mathbf{d}||$ as the initial state is prepared to lie near $\mathcal{W}^\mathbf{n}_N$ the SLSQP optimizer maintains a low baseline value of $\sim10^{-7}$ for $||\mathbf{d}||$ but with isolated spikes in between where it assumes larger values signifying departures of the trial wavefunction from  $\mathcal{W}^\mathbf{n}_N$ at some instances. Therefore our approach for finding $F_\mathrm{LL}(n_1)$ is not a strict constrained-search in
the sense of Levy~\cite{Levy1979, Mori-Sanchez2018} but since $||\mathbf{d}||$ is always small at convergence we ensure that the optimal state lies numerically close to $\mathcal{W}^\mathbf{n}_N$ as required.

With a means of calculating $F_\mathrm{LL}(n_1)$ at hand, we then proceed to perform density functional simulations for a given external potential asymmetry $\Delta v$ to identify the dimer ground state. This involves a variational search over occupation-numbers as outlined in table~\ref{tab:dvs} combined with \textit{on the fly} calculation of $F_\mathrm{LL}(n_1)$ using DC-VQE and it's derivative $F_\mathrm{LL}^{\prime}(n_1)$ using finite differences. Thus the density variational minimizer repeatedly calls DC-VQE at occupations $\mathbf{n}$ it encounters along the optimization trajectory. To initialize the DFT simulation we solve the non-interacting problem with $U=0$ and use the resulting occupation numbers as the starting guess. One could also for instance employ an approximate exchange-correlation potential $v_{xc}$~\cite{Capelle2013, Carrascal2015} in the dimer Hamiltonian to provide an initial density guess that will in most cases be closer to the true ground state density than the non-interacting guess. For the given $\Delta v$ we also compute the regular VQE ground state energy and density as the benchmark result. In figure~\ref{fig:dvs}(b,c) we show convergence with respect to VQE of the DFT energy and site occupation $n_1$ as a function of the number of $F_\mathrm{LL}(n_1)$ evaluations which represent the most intensive part of the simulation. We set $\Delta v=1$ and show trajectories for $U=1$ and $U=5$. In this parameter regime the correlation energy is typically sizable (see figure~6 of reference~\cite{Carrascal2015}) while at the same time the non-interacting and exact densities are expected to differ significantly. We find that around 20-30 evaluations of $F_\mathrm{LL}(n_1)$ are needed to converge the DFT energy and occupations to within $10^{-4}$ and $10^{-3}$ of the respective VQE reference values.

To conclude this section we note that while Levy-Lieb constrained-search~\cite{Levy1979, Levy1982, Lieb1983, Levy1985, Zhao1993} clarifies the coarse-graining step involved in switching our description of quantum systems from wavefunctions to densities, DFT based on repeated real-time evaluations of the exact $F_\mathrm{LL}[\mathbf{n}]$ using explicit constraints would be inefficient relative to unconstrained minimization in wavefunction space. Furthermore constrained optimization of parameterized quantum circuits is generally non-convex~\cite{Bittel2021, Tilly2021, Schuld2021, Kubler2021} and gradient based optimizers may get stuck in one of many local minima and therefore one needs to employ additional heuristics to locate the global minimum of the quantity being optimized. In the dimer case investigated here, similar to the case of variational quantum deflation on small molecules as noted previously~\cite{Higgott2019}, randomly initializing different starting guesses for the UCC ansatz parameters proved sufficient to identify the correct optimum for $F_\mathrm{LL}(\mathbf{n})$. For larger problem sizes more sophisticated global optimization schemes will be needed. Ideally specialized density-preserving parameterized ansatze that efficiently explore a specified density sector of Hilbert space would be desirable.

\begin{figure*}[htbp]
\includegraphics[angle=0, width=\textwidth]{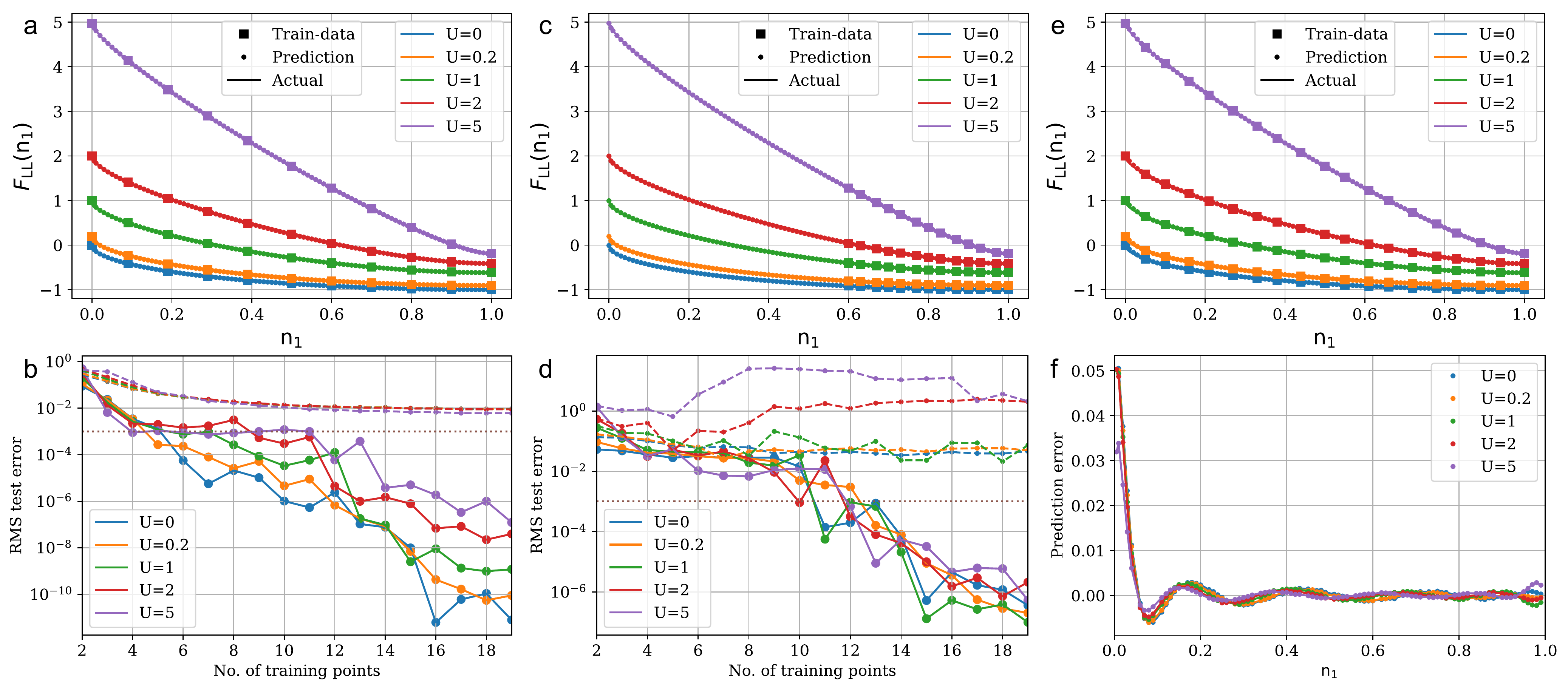}
\caption{\label{fig:llqk} (a) For the asymmetric Hubbard dimer, training, predicted and actual values of $F_\mathrm{LL}(n_1)$ when learning with the Levy-Lieb quantum kernel in an interpolaive setting. (b) Evolution of the RMS test error as a function of the number of training points for interpolative learning of $F_\mathrm{LL}(n_1)$ with different values of $U$. Results are shown both for the Levy-Lieb quantum kernel (filled-circles connect by solid lines) and the classical Gaussian kernel (dots connected by dashed lines). (c)  Training, predicted and actual values of $F_\mathrm{LL}(n_1)$ when learning with the Levy-Lieb quantum kernel in an extrapolative setting. (d) Evolution of the RMS test error as a function of the number of training points for extrapolative learning of $F_\mathrm{LL}(n_1)$ with different values of $U$. Results are shown both for the Levy-Lieb quantum kernel (filled-circles connected by solid lines) and the classical Gaussian kernel (dots connected by dashed lines). (e) Training, predicted and actual values of $F_\mathrm{LL}(n_1)$ when learning with the classical Gaussian kernel in an interpolative setting. (f) Distribution of test errors along the domain of $n_1$ when learning $F_\mathrm{LL}(n_1)$ in an interpolaive task with the classical Gaussian kernel.}
\end{figure*}
\section{The Levy-Lieb Quantum Kernel}\label{sec:llqk}
Restricting ourselves to non-degenerate settings where $n$ uniquely determines $\Psi$~\cite{Capelle2007} as outlined in section~\ref{sec:levylieb}, we can think of the Levy-Lieb mapping as implementing a particular \textit{feature map}~\cite{Schuld2019, Havlicek2019, Kubler2021, Schuld2021} encoding densities into the feature space $\mathcal{F}_\mathrm{LL}$ of pure-state density matrices $\rho_\mathrm{LL}[n] \equiv |\Psi_\mathrm{LL}[n]\rangle\langle\Psi_\mathrm{LL}[n]| \in \mathcal{F}_\mathrm{LL}$.
Additionally for the same density $n$, different embeddings can be realized through different choices for the operator sum $\hat{T} + \hat{W}$. So for a specified $\hat{T} + \hat{W}$, this allows us to define a \textit{fidelity} based Levy-Lieb quantum kernel (LLQK)
\begin{equation}
\kappa_\mathrm{LL}(n, n^{\prime}) \equiv \mathrm{Tr}\{\rho_\mathrm{LL}[n^{\prime}]\rho_\mathrm{LL}[n]\} = |\langle\Psi_\mathrm{LL}[n^{\prime}]|\Psi_\mathrm{LL}[n]\rangle|^2
\end{equation}
for densities $n, n^{\prime} \in \mathcal{X}_N \subset \mathcal{J}_N$ where $\mathcal{X}_N \equiv \{ n \in \mathcal{J}_N | n \iff \Psi_\mathrm{LL} \in \mathcal{W}^n_N  \}$ is the set on which the Levy-Lieb map is one-to-one. Note that this restriction is not needed if one only wishes to learn the functional $F_\mathrm{LL}[\mathbf{n}]$ alone and furthermore in situations with degeneracies one may optionally provide additional quantum numbers to pick a particular state on the degenerate manifold to uniquely specify the LLQK.  We can use the LLQK in a machine learning model of the form
\begin{eqnarray}
f_\mathcal{O}[n] = \mathrm{Tr}\{\rho_\mathrm{LL}[n]\mathcal{O}\} & = & \sum\limits_{i=1}^M \alpha_m \mathrm{Tr}\{\rho_\mathrm{LL}[n]\rho_\mathrm{LL}[n_m]\} \nonumber \\
& = & \sum\limits_{i=1}^M \alpha_m \kappa_\mathrm{LL}(n_m, n)
\end{eqnarray} where the goal is to \textit{learn} the optimal operator measurement $\mathcal{O}$ or equivalently the coefficients $\alpha_m \in \mathbb{R}$ given a set of $M$ labelled training data $\{n_m, f_\mathcal{O}[n_m]\}$. In particular since ground state density functionals are generated as expectation values of the form
\begin{equation}
O[n] = \langle \Psi_\mathrm{LL}[n]|\hat{O}|\Psi_\mathrm{LL}[n] \rangle = \mathrm{Tr}\{\rho_\mathrm{LL}[n]\hat{O}\}, 
\end{equation}
where \mbox{$\hat{O}=\hat{O}^{\dagger}$} is Hermitian, then given data \mbox{$(n_m, O[n_m])$}, \mbox{$m =1...M$} the optimal measurement $\mathcal{O}$ is just the projection of $\hat{O}$ in  $\mathcal{S}^M_\mathrm{LL} =\mathrm{span}\{\rho_\mathrm{LL}[n_1],\rho_\mathrm{LL}[n_2]...\rho_\mathrm{LL}[n_M]\}$. In other words, the reproducing kernel hilbert space (RKHS) of the LLQK~\cite{Schuld2021, Kubler2021} consists only of density functionals of the form $O[n] = \mathrm{Tr}\{\rho_\mathrm{LL}[n]\hat{O}\}$.

Here we explore the learning abilities of the LLQK in the context of the asymmetric Hubbard dimer. We calculate the LLQK for the dimer using the optimal UCC ansatz states obtained from DC-VQE for specified occupations $n_1$ in the relevant interval $\left(0,1\right]$. Since we consider pure state embeddings the fidelities are straightforwardly computed on a statevector simulator using unitary adjoints as $|\langle \Psi(\theta^{\prime} ) | \Psi(\theta)\rangle |^2 = |\langle 0 | \mathcal{U}^{\dagger}(\theta^{\prime} )\mathcal{U}(\theta)|0\rangle |^2$. On an actual quantum computer, the kernel entries can be estimated by evolving the initial state $|0\rangle$ with $\mathcal{U}^{\dagger}(\theta^{\prime} )\mathcal{U}(\theta)$ and calculating the ratio of all-zero outcomes to the total number of shots. We expect state fidelities to be robust to potential non-uniqueness of the mapping from densities $n_1$ to raw UCC parameters $\theta$ as under the Levy-Lieb procedure distances in density space should get mapped to physically meaningful distances in state space. Furthermore, since in this study we fix $2t=1$, we only vary the interaction term $\hat{W}$ by varying $U$ in the dimer Hamiltonian. $U$ can therefore be thought of as a hyperparamter that fixes the kernel and we study kernels $\kappa^{(U)}_\mathrm{LL}$ labeled by the $U$ value employed to generate the state embeddings leading to the kernel.  
\begin{figure*}[htbp]
\includegraphics[angle=0, width=\textwidth]{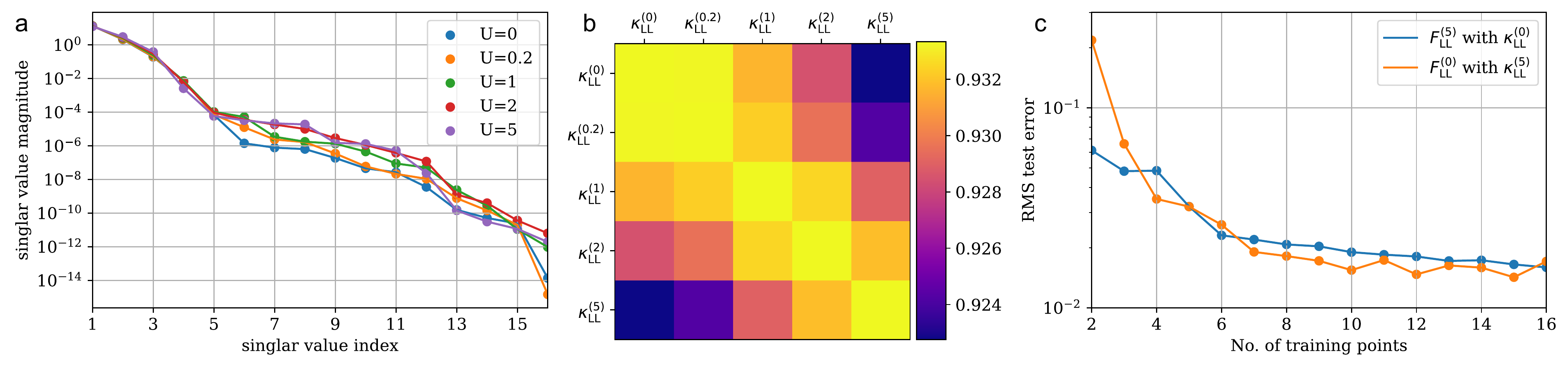}
\caption{\label{fig:svd} (a) Decay of the singular values of the Levy-Lieb quantum kernel matrices for different $U$. (b) Relative kernel alignment between Levy-Lieb quantum kernel matrices for different values of $U$. (c) RMS test errors while learning the interacting $F_\mathrm{LL}^{(5)}(n_1)$ with $U=5$ using the non-interacting quantum kernel $\kappa^{(0)}_\mathrm{LL}$ and vice versa.}
\end{figure*}

For a first set of learning tasks we consider $F_\mathrm{LL}(n_1)$ as the target function and construct a data set by sampling $n_1 \in (0,1]$ on a uniform grid with spacing 0.01 and at each $n_1$ calculate $F_\mathrm{LL}^{(U)}(n_1)$ and $\Psi^{(U)}(n_1)$ for different $U$. We then consider two kinds of train-test data splits: (L1) an \textit{in-distribution} or \textit{interpolative} learning task and (L2) an \textit{out-of-distribution} or \textit{extrapolative} learning task. For L1 we select $M$ training samples distributed as uniformly as possible over the interval $(0, 1]$ while including the end points $n_1 =0^+$ and $n_1 =1$ in the training set. All remaining points are used as test data points. This ensures that the sample mean for $n_1$ over the training and test sets is very close and in this 1D example every test data point has at at least one training point on either side. For L2 we select $M$ training samples distributed uniformly over the last $40\%$ of the interval $(0, 1]$ and include $n_1 = 1$ in the training set, while all other points are used as test data. In this instance the sample mean of $n_1$ over the training and test data is far apart and over 50\% of the test data can be reached only through 1D extrapolation. For a given $M$ and specified $U$, we precompute the kernel matrices $\kappa^{(U)}_\mathrm{LL}$ for the training and test data sets by evaluating fidelities on a noiseless quantum simulator. For each training set size $M$ considered, we verify the $M \times M$ $\kappa^{(U)}_\mathrm{LL}$ matrices are positive semidefinite and then use classical kernel ridge regression (KRR) as implemented in the scikit-learn library \cite{scikit-learn} to perform fitting and prediction of $F_\mathrm{LL}^{(U)}(n_1)$. The regularization hyperparameter of KRR is set to a small value of $4 \times 10^{-15}$ only to ensure numerical stability of the KRR fit process as $M$ approaches the rank of the kernel matrices. Thus we work in a setting where low training fit error is favored. For tasks L1, L2 we also use a classical Gaussian kernel $\kappa_G = \mathrm{exp}(-||\mathbf{n} - \mathbf{n}^{\prime}||^2/2\sigma^2)$ to fit and predict $F_\mathrm{LL}^{(U)}(n_1)$ for each U. In order to choose the Gaussian kernel hyperparameter $\sigma$ we use kernal alignment~\cite{Hubregtsen2021,Cristianini2001,Kandola2002} defined for two given kernel matrices $K, K^{\prime}$ as:
\begin{equation}\label{eq:kka}
A(K,K^{\prime}) = \frac{\langle K, K^{\prime} \rangle_F}{\sqrt{\langle K, K \rangle_F \langle K^{\prime}, K^{\prime} \rangle_F}} 
\end{equation}
Accordingly for each $(U,M)$, we separately optimize $\sigma$ so as to maximize $A(\kappa_G, \kappa^{(U)}_\mathrm{LL})$ on the training data set. 

Results for learning tasks L1, L2 as a function of the number of training points $M$ are shown in fig.~\ref{fig:llqk}. We use the root mean square (RMS) error on the test data set as our performance metric. Figure~\ref{fig:llqk}(a), shows plots for $F_\mathrm{LL}(n_1)$ with $M$=11 training points and corresponding predicted values from task L1 overlaid. The training data points occur at roughly regular intervals on $(0, 1]$ and the predictions are indistinguishable from reference data to the naked eye. In fact at $M=11$ the RMS test error for all values of $U$ is under $10^{-3}$ as seen from figure~\ref{fig:llqk}(b) where the evolution of the test error is plotted against $M$. We see that with the LLQK the test error quickly falls to under $10^{-2}$ at $M=4$ for all of the $U$ values considered and then continues to fall gradually as more training points are added eventually reaching lower than $10^{-6}$. The RMS test errors obtained by using the Gaussian kernel on task L1 are also plotted in figure~\ref{fig:llqk}(b) and show a saturation of the prediction error at around $10^{-2}$ over the range of $M$ investigated. We plot predicted values for $F_\mathrm{LL}(n_1)$ obtained from the Gaussian kernel within task L1 for $M=19$ in figure~\ref{fig:llqk}(e). Since the test error is around $10^{-2}$ the fit looks good to the naked eye. In order to understand why the error does not seem to improve beyond $10^{-2}$ we plot the error distribution over the entire test data set on the $n_1$ interval $(0,1)$ in figure~\ref{fig:llqk}(f). We see that the prediction errors are primarily concentrated near $n_1 \rightarrow 0$ where $F_\mathrm{LL}(n_1)$ exhibits a sharp increase in its slope~\cite{Carrascal2015} and to a lesser extent near $n_1 \rightarrow 1$. Thus within the chosen setup for task L1, the Gaussian kernel does not generalize as well near $n_1 \rightarrow 0$ as it is able to elsewhere. Further, on the interpolative learning task L1, we do not find significant improvements in the prediction error of the Gaussian kernel either by increasing the KRR regularization parameter and allowing for higher training loss or by manually tuning $\sigma$. However, since the Gaussian kernel is a universal kernel~\cite{Micchelli2006}, we expect to be able to reduce the prediction errors by providing more training data near $n_1 \rightarrow 0$.

In fig~\ref{fig:llqk}(c) we plot training and predicted data points as obtained with the LLQK for $F_\mathrm{LL}(n_1)$ within task L2 where the training set is drawn only from the last $40\%$ of the $(0,1]$ interval. We plot the fit obtained for $M=13$ as at this training set size the LLQK RMS test error is under $10^{-3}$ for all U values considered. Figure~\ref{fig:llqk}(d) shows the evolution of the LLQK prediction error as a function of $M$ for task L2. We see that in extrapolating $F_\mathrm{LL}(n_1)$ out to densities far away from the training interval the error falls more gradually compared to task L1 but once a sufficient number of training points are provided it declines further to around $10^{-6}$. Thus in this simple model of the asymmetric Hubbard dimer, the LLQK is able to generalize with high accuracy regardless of how the training and test data are distributed over the domain of $n_1$. In contrast, the Gaussian kernel is not expected to generalize with naive KRR to extrapolative test data and especially with the small regularization hyperparameter originally chosen, we see from Figure~\ref{fig:llqk}(d) that prediction errors are unacceptably high. We find (not shown) that increasing the KRR regularization parameter on task L2 for the Gaussian kernel from $4\times10^{-15}$ to around $10^{-5}$ improves prediction errors somewhat from $10^0$ to around $10^{-1}$ at which level they remain as a function of $M$. 

The observed behavior of the LLQK for the dimer suggests it leads to a restricted model with a low effective dimension in feature space. For $M=16$ we show the decay of the singular values of the $\kappa^{(U)}_\mathrm{LL}$ matrices for different $U$ in figure~\ref{fig:svd}(a). It is apparent that the magnitudes of the singular values initially decay rapidly suggesting a low effective dimension. Additionally we find that as the size of the training set $M$ increases, the numerically computed rank of the $\kappa^{(U)}_\mathrm{LL}$ matrices stops growing around 16 - 21 irrespective of how the training data is distributed over the domain of $n_1$. So even as  nominally the 4-qubit system is associated with a 256 dimensional feature space, the Levy-Lieb embedding for the dimer is itself restricted to a small subspace and once enough training points are provided to effectively span important features within this subspace, operator measurements fit to the training data are predictive for test data if the function being predicted is of the form $O^{(U)}[n_m] = \mathrm{Tr}\{\rho_\mathrm{LL}^{(U)}[n_m]\hat{O}\}$. To emphasise this latter point we consider the transferability of the $\kappa^{(U)}_\mathrm{LL}$ kernels for a specified $U$ in terms of learning density functionals associated with a different interaction strength $U^{\prime}$. To this end, firstly in figure~\ref{fig:svd}(b), we show the relative kernel alignment computed using equation~\ref{eq:kka}, between $\kappa^{(U)}_\mathrm{LL}$ matrices associated with different $U$. Not surprisingly the alignment $A(\kappa^{(U^{\prime})}_\mathrm{LL}, \kappa^{(U)}_\mathrm{LL})$  decreases as $|U-U^{\prime}|$ increases. In figure~\ref{fig:svd}(c) we show the prediction errors associated with an interpolative learning task where we attempt to learn the interacting Levy-Lieb functional $F_\mathrm{LL}^{(5)}(n_1)$ using the non-interacting LLQK $\kappa^{(0)}_\mathrm{LL}$ and vice versa. We find in this case that the prediction errors level out just above $10^{-2}$ as the number of training data points approaches the embedding rank, which is qualitatively different behavior than what we observed for each kernel $\kappa^{(U)}_\mathrm{LL}$ when it was used to learn functionals for the same $U$. 
\begin{figure}[htbp]
\includegraphics[angle=0, width=0.5\textwidth]{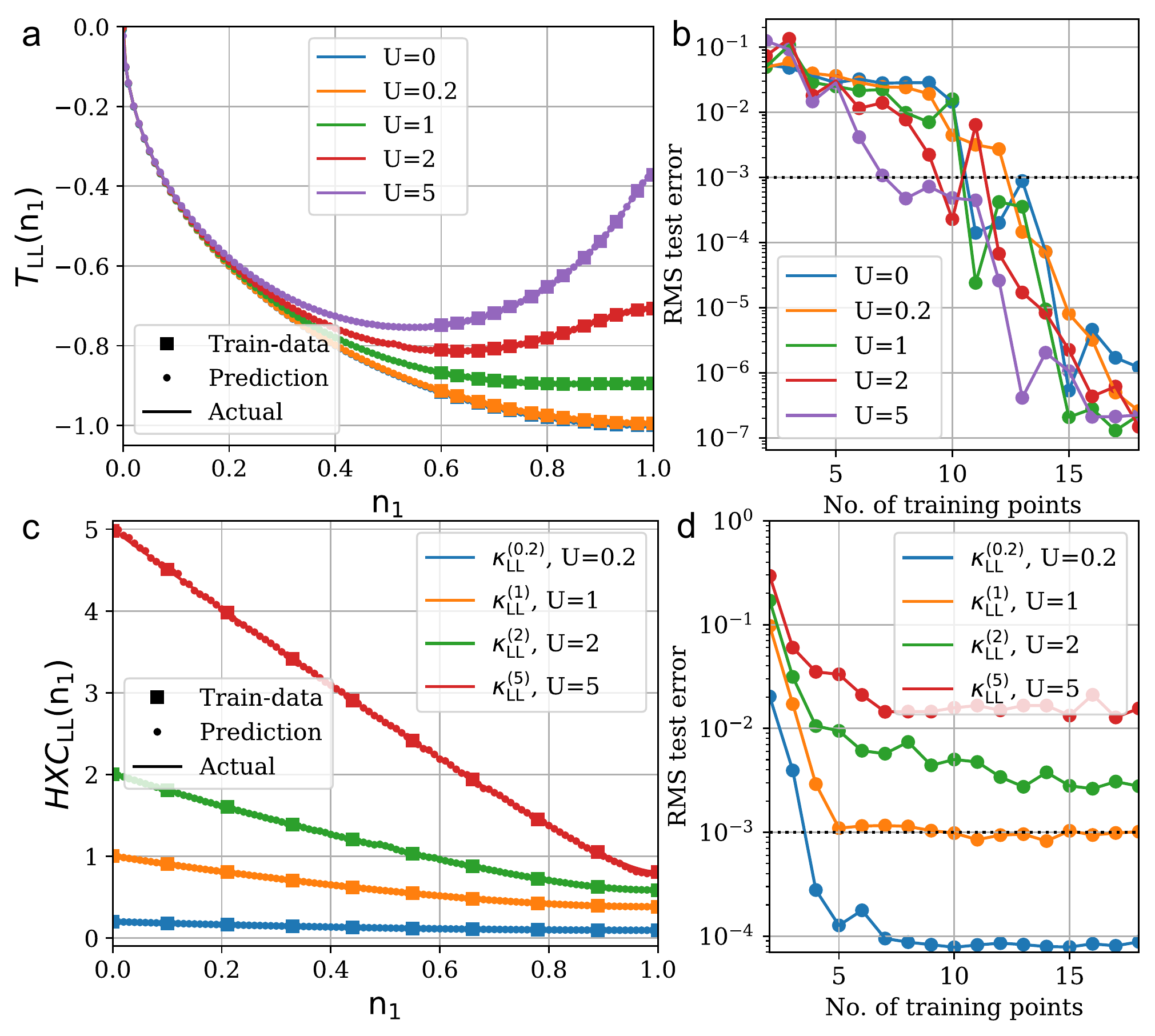}
\caption{\label{fig:thxc}(a) Training, predicted and actual values of $T_\mathrm{LL}(n_1)$ when learning with the Levy-Lieb quantum kernel in an extrapolative setting. (b) Evolution of the RMS test error as a function of the number of training points for extrapolative learning of $T_\mathrm{LL}(n_1)$ using the Levy-Lieb quantum kernel (filled-circles connect by solid lines) for different values of $U$ (c)  Training, predicted and actual values of $HXC_\mathrm{LL}(n_1)$ when learning with the Levy-Lieb quantum kernel in an interpolative setting. (d) Evolution of the RMS prediction error as a function of the number of training points for interpolative learning of $HXC_\mathrm{LL}(n_1)$ for different values of $U$.}
\end{figure}

Next, we show that results obtained in terms of learning $F_\mathrm{LL}(n_1)$ are not specific to its particular form and other density functionals are expected to behave similarly. Accordingly we consider two additional functionals the first being the kinetic energy functional
\begin{equation}
T^{(U)}_\mathrm{LL}[n] = \langle \Psi_\mathrm{LL}^{(U)}[n]|\hat{T}|\Psi_\mathrm{LL}^{(U)}[n] \rangle
\end{equation}
and the second being the so-called Hartree-Exchange-Correlation ($HXC$) functional defined as
\begin{equation}
\begin{aligned}
HXC^{(U)}_\mathrm{LL}[n] &= \langle \Psi_\mathrm{LL}^{(U)}[n]|\hat{T} + \hat{W}|\Psi_\mathrm{LL}^{(U)}[n] \rangle \\
&- \langle \Psi_\mathrm{LL}^{(0)}[n]|\hat{T}|\Psi_\mathrm{LL}^{(0)}[n] \rangle \\
&= F_\mathrm{LL}^{(U)}[n] - F_\mathrm{LL}^{(0)}[n]
\end{aligned}
\end{equation}

The kinetic energy functional $T^{(U)}_\mathrm{LL}[n]$ is interesting because even for the dimer, it is a non-monotonic function and its form evolves non-trivially with $U$ (see Fig.~\ref{fig:thxc}(a)). The $HXC^{(U)}_\mathrm{LL}[n]$ functional is instructive because for the dimer as seen graphically in figure~\ref{fig:thxc}(c) it has a relatively simple monotonic behavior as a function of $n_1$ but in its definition it involves both the interacting $\Psi_\mathrm{LL}^{(U)}[n]$ and non-interacting $\Psi_\mathrm{LL}^{(0)}[n]$ states. We consider learning $T^{(U)}_\mathrm{LL}(n_1)$ for the dimer in an extrapolative setting similar to task L2 described previously and find as shown in figure~\ref{fig:thxc}(a,b) that in spite of its non-monotonic behavior, $T^{(U)}_\mathrm{LL}(n_1)$ is easily generalized by $\kappa^{(U)}_\mathrm{LL}$ to far away test data with high accuracy as the number of training data points approaches the kernel rank. For $HXC^{(U)}_\mathrm{LL}(n_1)$ we consider an interpolative setting similar to task L1 above and attempt to learn $HXC^{(U)}_\mathrm{LL}(n_1)$ using $\kappa^{(U)}_\mathrm{LL}$. We also increase the KRR regularization parameter in this instance to $10^{-6}$ to improve generalization behavior. Note that $HXC^{(0)}_\mathrm{LL}(n_1) = 0$ for the dimer by definition and is ignored. We find once again that because of the involvement of states with $U=0$, kernels $\kappa^{(U)}_\mathrm{LL}$ with $U>0$ show a prediction error profile for $HXC^{(U)}_\mathrm{LL}(n_1)$ that flattens out as a function of $M$ (see Fig.~\ref{fig:thxc}(d)) in sharp contrast to the generalization ability apparent for $T^{(U)}_\mathrm{LL}(n_1)$. Furthermore we see that the level at which the error stops improving depends on $U$. This illustrates the selective nature of the RKHS associated with the Levy-Lieb embedding of densities.

Our results for the Hubbard dimer suggest that if somehow we only had access to the Levy-Lieb quantum kernel but not the states themselves, we could still calculate density functionals of observables to high accuracy given sufficient training data. Since the dimer is a small system, assuming we have the correct kernel at hand say in the form of the LLQK, it is easy to provide enough training data to exceed the effective dimension of the embedding and achieve very low prediction error. As we do not conduct system size dependent studies in this work, we do not draw conclusions about the scaling of the effective dimension associated with the Levy-Lieb embedding and thus data requirements for learning density functionals. In general such an analysis would have to be conducted in a context specific manner as based on complexity theoretic arguments~\cite{Baker2020, Schuch2009}, we do not expect generic quantum advantage for machine learning the universal functional of DFT. The learning abilities of general quantum models of the form $f(x) = \mathrm{Tr}(\hat{O}\rho(x))$ have been analyzed recently by several authors~\cite{Schuch2009, Schuld2019, Schuld2021, Huang2021, Kubler2021}. Huang et al showed that while quantum kernels of the type $\kappa_Q(x, x^{\prime}) = |\langle x|x^{\prime}\rangle|^2 = \mathrm{Tr}(\rho(x)\rho(x^{\prime}))$ can be expected to learn a very general class of functions, in the worst case, the training data requirements to achieve small prediction errors could be very large.  Furthermore recent theoretical works~\cite{Huang2021, Kubler2021} have also discussed the intrinsic weakness of fidelity based kernels with regards to exponentially decaying off-diagonal kernel matrix elements in high dimensions and suggested projected kernels based on reduced density matrices (RDMs). Fortunately for most observables relevant to DFT and materials science mapping densities to feature spaces based on RDMs would also suffice~\cite{Ludena2013}.  Additionally with regards to classical machine learning, it should be noted that techniques related to machine learning DFT~\cite{Li2016, nelson2019machine, Moreno2020, vonLilienfeld2020, Kirkpatrick2021} employing both implicit kernel methods and deep learning are now very mature and results from the Gaussian kernel based KRR used in this work for illustration purposes on the asymmetric Hubbard dimer are not meant to suggest quantum advantage for DFT.\\

\section{Conclusions}\label{sec: summary}
We discussed the constrained-search formulation of density functional theory (DFT) within the context of near-term quantum algorithms highlighting the relationship between exact density functionals and underlying wavefunctions. We used parameterized quantum circuits to implement a form of density variational minimization involving run-time calculations of the exact Levy-Lieb functional and illustrated its equivalence to unconstrained wavefunction optimization for obtaining ground states. Interpreting the Levy-Lieb mapping from densities to wavefunctions as a feature space embedding into pure states we demonstrated a quantum kernel that allows us to learn density functionals of observables without obscuring the underlying state space. We hope that our work contributes to improving the explainability of DFT and other reduced variable theories to a broader audience interested in quantum theory.

\bibliography{levylieb, amol, Qiskit}

\end{document}